\documentclass{article}
\usepackage{graphicx}

\title{Layered Heaps Beating Standard and Fibonacci Heaps in Practice}
\author{P.~Huggins}

\usepackage{amsmath,amsthm}

\begin{document}

\maketitle

\abstract{We consider the classic problem of designing heaps. 
Standard binary heaps run faster in practice than Fibonacci heaps but have 
worse time guarantees. Here we present a new type of heap that runs faster in
practice than both standard binary and Fibonacci heaps, but has asymptotic 
insert times arbitrarily better than $O(\log n)$, namely
 $O((\log n)^{1/m})$ for arbitrary positive integer $m$. 
Our heap is defined recursively and maximum run time speed up occurs when a recursion depth of $1$ is used, i.e. a heap of heaps.}

\section{Layered Heaps}

We will define $M$-layered heaps for arbitrary integer $M \geq 1$. For $M = 1$
the $M$-ary heap is a standard binary heap stored in an array. For $M \geq 2$, a $k_M$-ary heap is used, with $k_M = 2^{(\log n)^{k_{M-1}*(M-1)/M}}$. Then by inductive hypothesis, insert operations on the children heaps will be $(\log n)^{(M-1/M)*(1/(M-1))} = (\log n)^{1/M}$. Furthermore the height of the $M$-ary heap will also be $(\log n)^{1/M}$. So insert operations take $O((\log n)^{1/M})$ time for any $M$ we want. Pop/delete functions take standard $O(\log n)$ time, because we may need to do a children heap operation which takes $(\log n)^{(M-1)/M}$ time a total of $(\log n)^{1/M}$ times, i.e. the height of the $M$-ary layered heap.

The operations and running times for them are explained in the following pseudocode:

\vskip 0.2in

{\bf INSERT INTO M-ARY HEAP $A$:} Before swapping, start by placing the element at end of array (position $n = N$). Then do the following:
\begin{itemize}
\item Set $k = 2^{(\log N)^{(M-1)/M}}$
\item While $n > 0$:
\begin{itemize}
\item If $A[n] > A[n/k]$ then swap their values
\item Else insert $A[n]$ into children $(M-1)$-ary layered heap that contains position $n$ in the array. (Recursive) Then BREAK.
\item Set $n := n/k$
\end{itemize}
\end{itemize}

\vskip 0.2in 
{\bf POP OUT OF M-ARY HEAP $A$}:
\begin{itemize}

\item Set $k = 2^{(\log N)^{(M-1)/M}}$
\item Store and remove root element of $M$-ary heap.
\item Put the last item in the heap at the root.
\item Swap downwards with top of children heap while top of children heap is greater than element. (Recursively balance the $(M-1)$ary heap in time $O((\log n)^{(M-1)/M})$).
\item Break when element is greater than top of current children heap.
\item Return popped top of heap

\end{itemize} 

\vskip 0.2in

As can be seen, the running time for insert is $O((((\log n)^{(M-1)/M}))^{1/(M-1)})$ which is $O((\log n)^{1/M}$. The running time for pop is $O((\log n)^{(M-1)/M + 1/M} = O(\log n)$.

\section{Popular Competing Heaps}
In [1] the Fibonacci heap is presented, which has (amortized) constant insert time, and standard $O(\log n)$ delete/pop time. The amortized running times were later improved to strict running time bounds per operation in a later publication. However, in practice, the constants associated with various Fibonacci heaps are too large to outperform a standard binary tree. Thus, due to its simplicity and faster running time, binary heaps are traditionally what is taught and used.

\section{Running time comparisons for insert/pop}

To simulate situations where asymptotically faster insertions in heaps may be better than traditional heaps, we did a 10 to 1 simulation where 1000 elements would be added and then 100 elements would be popped, where the $i$th insert inserted the value $i$ (and the heap is a max-heap), and repeated over and over with running times being recorded as a function of the size of the heap. 
Binary heaps are faster than Fibonacci heaps for practical data sizes 
in practice. Furthermore, analysis of our recursively defined $M$-ary 
layered heaps made it clear that the constants become too large to be 
overcome in practice unless $M=2$. Thus we compared the 2-layer heap to 
the traditional binary heap. Results are shown in the figure, where $N$ is the 
number of elements in the heap as the heap grows. Results were computed out to $N = 2^{32}$ and then extrapolated to $N = 2^{40}$ to cover all feasible data sizes.

\begin{figure}

\begin{centering}

\includegraphics[width=4in]{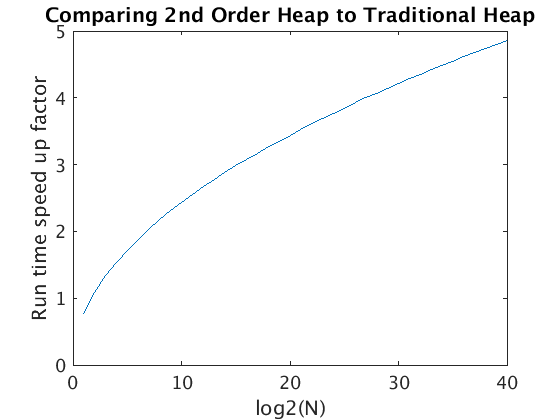}

\end{centering}

\end{figure}

As the figure shows, despite the $O(\log n)$ time for both heaps when a pop is performed, the 2-ary layered heap has good cache performance when processing a children heap because it only has about 50 elements and so usually they all fit into cache after one memory access into the children heap. Thus the running time in practice for inserts into the 2-ary layered heap look more like an inflated $O(\sqrt{\log n})$. Memory use is identical for both heaps.

\section{Discussion}

Although $M$-ary layered heaps are interesting from a theoretical point of 
view for arbitrary $M$, giving asymptotic insert running time arbitrarily 
closer and closer to constant, in practice the 2-ary layered heap
is the fastest in practice and can run up to 3-4 times faster than a binary
heap for reasonable data sizes.

In fact, 2-ary heaps are easy enough to describe and implement and analyze 
directly (as opposed to using induction/recursion for $M > 2$), they should
probably be taught in data structures courses after standard binary heaps
are presented.

\section{Bibliography}

\begin{enumerate}

\item Fredman, Michael Lawrence and Tarjan, Robert E. (1987). "Fibonacci heaps and their uses in improved network optimization algorithms".  Journal of the Association for Computing Machinery 34 (3): 596–615. doi:10.1145/28869.28874.

\end{enumerate}

\end{document}